\documentclass[a4paper,11pt]{article}
\pdfoutput=1 

\usepackage{jheppub} 

\usepackage[T1]{fontenc} 

\usepackage{amsmath}
\usepackage{bbm}

\usepackage{slashed}
\usepackage{subcaption}
\newcommand\myworries[1]{\textcolor{red}{#1}}

\title{\boldmath Solutions to the Dirac equation in the background of 1 monopole with 2 singularities }

\author{Thomas Harris}

\affiliation{Department of Mathematics, University of Arizona, \\617 N Santa Rita Ave, Tucson 85716, USA}

\emailAdd{thomasharris@math.arizona.edu}

\abstract{We use the Nahm transform to construct explicit $L^2$ solutions to the Dirac equation in $\mathbb{R}^3$ in the background of one nonabelian $U(2)$ monopole with one positive and one negative Dirac singularity. }

\begin{document} 
\maketitle
\flushbottom

\section{Introduction} 

In 1931, Dirac argued \cite{Dirac} that the existence of a magnetic monopole would imply charge quantization of electrons. $U(1)$ electromagnetism prohibits monopoles via the Bianchi identity:

\begin{equation}
\label{maxwellhomo}
\rho_{\textrm{mag}} = \nabla \cdot B = 0.
\end{equation}

But by working in a $U(2)$ Yang-Mills-Higgs theory, solitons which carry magnetic charge such as the 't Hooft-Polyakov monopole \cite{tHooft} can arise. BPS monopoles are then considered solutions to the Bogomolny equations

\begin{equation}
F_A = *D_A \Phi,
\end{equation}

where $A$ is the connection of some $U(2)$ bundle over $\mathbb{R}^3$, $\Phi \in \mathfrak{u}(2)$, $D_A$ is the covariant derivative and $F_A$ is the curvature. The magnetic charge is 

\begin{equation}
\lim\limits_{R \rightarrow \infty} \int_{S^2_R} \mathrm{Tr}(\Phi F_A). 
\end{equation}

In general, these are non-linear partial differential equations. Nahm \cite{Nahm1} adapted the ADHM construction of instantons to obtain static self dual solutions to the self-dual Yang-Mills equations:

\begin{equation}
F_A = *F_A.
\end{equation}

By considering the time component of this static connection as the Higgs field, we can recover the Bogomolny equations. Following Nahm, we consider a family of solutions to the Bogomolny equations given by modifying $\Phi$ to $\Phi - t \mathbbm{1}$ where $t$ is some constant. The zero modes of the Dirac equation induce a vector bundle over some interval $t \in \mathcal{I} \subset \mathbb{R}$ with induced connection $\nabla$ and endomorphisms $T_1, T_2, T_3$ solving the Nahm equations \eqref{Nahmequation}. An analogous process, the Nahm transform, can be applied to the data $(\nabla, T_1, T_2, T_3)$ over $\mathcal{I}$ to recover the original connection.

In this paper, we use the Nahm transform to construct solution to Dirac equation. We refer the reader \cite{Jardim} for a survey of the Nahm transform.

\subsection{Bogomolny Equations}

Given compact lie group $G$ and a principal $G$-bundle over $\mathbb{R}^3$ with connection $A$ and Higgs field $\Phi$, we can define a Lagrangian density via

\begin{equation}
\mathcal{L} = \frac{1}{2}Tr(F_A \wedge *F_A) + Tr(D_A \Phi \wedge *D_A \Phi).
\end{equation}

 Here $F_A$ denotes the curvature and $D_A$ is the covariant derivative. The resulting energy from this Lagrangian is
 
 \begin{equation}
 E = \int _{\mathbb{R}^3} \frac{1}{2}Tr(F_A \wedge *F_A) + Tr(D_A \Phi \wedge *D_A \Phi).
 \end{equation}
 
We are looking for pairs $(\Phi, A)$ which minimize $E$. Under gauge transformation $g$, $\Phi,A,F_A$ transform as follows:

\begin{equation}
\Phi \rightarrow g^{-1} \Phi g
\end{equation}

\begin{equation}
A \rightarrow g^{-1} A g - g^{-1} dg 
\end{equation}

\begin{equation}
F_A \rightarrow g^{-1} F_A g.
\end{equation}

Note that while $\mathcal{L}$ is invariant under gauge transformation, $\Phi$ need not be. Denote the stabilizer of $\Phi$ by $H \le G$. We embed a 2-sphere of radius $R$, $S^2_{R}$, into $\mathbb{R}^3$. By looking at very large $R$, $\Phi$ induces a map $S^2_{R} \rightarrow G/H$. In the case that $G/H$ is a $2-$sphere, this map divides our monopoles into topological classes determined by the degree of said map, which we will call $Q$. You could ask, given such a $Q$, how could we find a pair $(\Phi, A)$ of the topological class? The lower bound of the energy $E$ is proportional to $Q$. By extremizing the energy, we get the Bogomolny equations: 

\begin{equation}
F_A = *D_A \Phi.
\end{equation}

\subsection{Monopoles}

Two of the Maxwell's equations are

\begin{equation}
\label{maxwellhomo}
\rho_{\textrm{mag}} = \nabla \cdot B = 0
\end{equation} 

\begin{equation}
\label{maxwellinhomo}
\nabla \cdot E = 4\pi  \rho _e,
\end{equation}

where $\rho_{\textrm{mag}} $ represents magnetic charge density, $B$ the magnetic field and $E$ the electric field. Dirac considered the magnetic field that would arise from a point-like magnetic particle

\begin{equation}
\label{maxwellmonopole}
 B = m \frac{\overrightarrow{r}}{r^3},
\end{equation}

which led him to the Dirac monopole. A property of this solution is that it is defined on $\mathbb{R}^3$ minus a line, the Dirac string. We refer to the reader to \cite{solitons} for the connection between Dirac's monopole and the Bogomolny equations. 

The Dirac monopole is now interpreted as a $U(1)$ monopole and the Dirac string a gauge artifact. Kronheimer \cite{Kron} also considered monopoles where $G$ is $U(1)$. In this case, the Bogomolny equations imply:

\begin{equation}
d*d \Phi = 0 .
\end{equation}

Thus $\Phi$ is harmonic. If $\Phi$ is bounded and positive, and defined on $\mathbb{R}^3$, then $\Phi$ is constant. Kronheimer then allows the possibility that the monopole has singular points. Singular monopoles on compact manifolds were analyzed in \cite{Pauly}. The 't Hooft-Polyakov monopole \cite{tHooft},\cite{Polyakov} is non-singular and contributes a $U(2)$ gauge field. In \cite{Durcan}, Durcan explicitly computes the Higgs field, connection, and solutions to the Dirac equation used in this paper. Solutions to Dirac equation for $N$ abelian BPS monopoles all located at the same place were found in \cite{Cheng}. Solutions for  dipoles were found in \cite{Baal} and solutions for $N$ abelian BPS monopoles at generic positions were found in \cite{Poirer}.

We can construct solutions to Bogomolny equations indirectly. If $E$ is a Hermitian vector bundle associated to our principal $G$ bundle, we can tensor it with the spinor bundle over $\mathbb{R}^3$ and define a Dirac operator on the resulting bundle. The Nahm Transform can be used to construct BPS monopoles, as well as solutions to the Dirac equation coupled to a BPS monopole \cite{Schenk}. In this note we construct explicit normalizable solutions to the Dirac equation over $\mathbb{R}^3$ coupled to gauge fields arising from one BPS monopole with two Dirac singularities with opposite charge. There are three specified locations on our manifold, and if we choose two of them to move to infinity, we can recover a known solution to the Dirac equation.

 This work is an extension of Brian Durcan's Master Thesis \cite{Durcan}. Durcan constructed  solutions to the Bogomolny equations by using the Nahm Transform. Durcan did this by solving the Nahm equations over $\mathcal{I}$, then finding normalized zero modes of the resulting Dirac equation and finally using the zero modes to construct the Higgs field and connection over $\mathbb{R}^3$ which solved Bogomolny equations.

\section{The Nahm Transform}

	We consider a Hermitian vector bundle $E$ with structure group $G$ over $\mathcal{I} := (\infty, -\lambda) \cup (-\lambda,\lambda) \cup (\lambda, \infty) $ parametrized by $s$.  Solutions to the Nahm equations \cite{Nahm1} are $T_0, T_1, T_2, T_3$, where $\frac{d}{ds} + i T_0$ is a connection on $E$ and  $T_1, T_2, T_3$ are endomorphisms of $E$ living in the adjoint representation of the structure group of $E$, satisfying the Nahm equations: 
	
	\begin{equation} \label{Nahmequation}
	\frac{dT_i}{ds} - \mathrm{i} [T_0, T_i] = - \mathrm{i}[T_j, T_k], 
	\end{equation} 
	
	with $(i,j,k)$ being any cyclic permutation of $(1,2,3)$. 

	By choice of gauge, we can set $T_0 = 0$ resulting in 
	
	\begin{equation}
	\label{gaugedNahm}
	\frac{dT_i}{ds} = -\mathrm{i}[T_j, T_k].
	\end{equation} 

	Let us briefly summarize the Nahm transform \cite{Nahm1}. Since there are two Dirac operators appearing in this paper, we will refer to the first one as the Weyl operator. We use the following notation
	
	\begin{equation}
	\slashed{x} = \sigma_1 x_1 + \sigma_2 x_2 + \sigma_3 x_3,
	\end{equation}
	where the Pauli matrices $\sigma_i$ are given by:
	
	\begin{equation}
	\sigma_1 = \begin{pmatrix}
	0 & 1 \\
	1 & 0
	\end{pmatrix}
	\quad
	\sigma_2 = \begin{pmatrix}
	0 & -i \\
	i & 0
	\end{pmatrix}
	\quad
	\sigma_3 = \begin{pmatrix}
	1 & 0 \\
	0 & -1
	\end{pmatrix}.
	\end{equation}
	Given a point $x \in \mathbb{R}^3$ the twisted Weyl operator is defined as
	\begin{equation}
	\label{eq:weyloperator}
	{\slashed{D}_x}=\frac{d}{{ds}} - (\slashed{T}(s) - \slashed{x}), 
	\end{equation}
	
	and its Hermitian conjugate
	\begin{equation} \label{weylUnmodified}
	{\slashed{D}^\dagger_x}=-\frac{d}{{ds}} - (\slashed{T}(s) - \slashed{x}) .
	\end{equation}
	
	The Weyl operator acts of sections of $ L^2 (\mathcal{I}, \mathbb{C}^2 \otimes E)$. The Dirac Laplacian ${\slashed{D}^\dagger_x}{\slashed{D}_x}$ is:
	\begin{equation}
	{\slashed{D}^\dagger_x}{\slashed{D}_x} 
	= \mathbbm{1} 
	( - \frac{d^2}{ds^2} + + \sum_{j = 1}^{3} T_j ^2 ) 
	- \mathrm{i} \sum_{i=1}^{3} \sigma_i (\frac{dT_i}{ds}  + \mathrm{i}[T_j, T_k]). 
	\end{equation}
	
	By \eqref{gaugedNahm}, the imaginary part vanishes, thus the Dirac Laplacian is real.

	Let $\rho_x$ be a matrix whose columns form an orthonormal basis for Ker $\slashed{D}^\dagger_x$ and $G_x(t,s)$ the Green's function of ${\slashed{D}^\dagger_x}{\slashed{D}_x}$. $\rho_x$ induces a frame of a Hermitian vector bundle over $\mathbb{R}^3$ with Higgs field $\Phi$ and connection $A_j$ defined as 

	\begin{equation}
	\Phi = \int s  \rho_x^{\dagger}  \rho_x  \, ds
	\end{equation}
	\begin{equation}
	A_j = i \int \rho_x^{\dagger} \frac{d}{{dx^j}}\rho_x  \, ds .
	\end{equation}
	
	This choice of $\Phi$, $A_j$ is a solution the Bogomolny equations. \cite{Nahm1}
	
	By tensoring the resulting Hermitian vector bundle with the spin bundle over $\mathbb{R}^3$, using $\Phi$ and $A_j$, we define the twisted Dirac operator over $\mathbb{R}^3$ as
	\begin{equation}
	D_t = -(\Phi -t) +   \overset{\rightharpoonup }{\sigma } \otimes (\frac{d}{d \overset{\rightharpoonup }{x}} - i \overset{\rightharpoonup }{A}).
	\end{equation}
	
	This is the monopole Dirac operator. The zero modes of the Dirac equation, $D_t\chi_t = 0$, are given by~\cite{Schenk}
	\begin{equation}
	 \widetilde{\chi}_t(x)  = \int \rho^\dagger_x(s)  G_x(t,s)  \, ds
	\end{equation}
	\begin{equation}
	\widetilde{\chi}_t = \chi^T_t \begin{pmatrix}
	0 & 1 \\
	-1 & 0
	\end{pmatrix}. 
	\end{equation}
	
\section{Solutions to the Weyl equation }

	If $E$ is rank 1, then $T_i$ is a function instead of a matrix valued function and solutions to the Nahm equations must then be constant, with possible jumps at $\pm \lambda$. We use the following notation for the solution to Nahm equations. 

\begin{equation}
(T_1(s), T_2(s), T_3(s)) = \overrightarrow{T}(s) =  \left\{
\begin{array}{ll}
\overrightarrow{T}_{D_1}  \in \mathbb{R}^3 & \quad s < -\lambda \\
\overrightarrow{T}_{'tHP} \in  \mathbb{R}^3 & \quad -\lambda < s < \lambda \\
\overrightarrow{T}_{D_2}  \in \mathbb{R}^3 & \quad s > \lambda
\end{array}
\right.
\end{equation}

Here we are identifying the $(T_1(s), T_2(s), T_3(s))$ with points in $\mathbb{R}^3$. $\overrightarrow{T}_{D_1},\overrightarrow{T}_{D_2}, \overrightarrow{T}_{'tHP}$ will be the locations of the negative Dirac singularity, the positive Dirac singularity and BPS monopoles respectively. 

To account for the jump discontinuities, the Hermitian conjugate of the Weyl operator is modified \cite{Cherkis} to

\begin{equation}
\label{eq:weyloperator}
{\slashed{D}_x}^{\dagger}
=-\mathbbm{1}\frac{d}{{ds}} - 
(\slashed{T}(s) - \slashed{x})  +\delta  (\lambda +s) Q_{1-}+\delta  (s-\lambda ) Q_{2+}. 
\end{equation}

The spinor $Q_{i \pm}$ satisfies

\begin{equation}
	Q_{i \pm}Q^\dagger_{i \pm} = | \overrightarrow{T}_{'tHP} - \overrightarrow{T}_{D_i}| \pm \overrightarrow{\sigma} \cdot (\overrightarrow{T}_{'tHP} - \overrightarrow{T}_{D_i}),
\end{equation}

where $\overrightarrow{\sigma} = (\sigma_1, \sigma_2, \sigma_3)$.

Let us introduce the following relative positions to simplify formula:

\begin{equation}
\overrightarrow{z_i} =\overrightarrow{x} -\overrightarrow{T}_{D_i}
\end{equation}
\begin{equation}
\overrightarrow{r} =\overrightarrow{x} - \overrightarrow{T}_{'tHP}
\end{equation}

Here $\overrightarrow{x}$ is a location in $\mathbb{R}^3$, so $\overrightarrow{z_i}$ represents the  location relative to one of the Dirac singularities and $\overrightarrow{r}$ represents the location relative to the BPS monopole. 

We also define $\alpha_1, \alpha_2, \theta$: 
\begin{equation}
e^{2 r \alpha _i}=\frac{z_i + |\overrightarrow{z_i} - \overrightarrow{r}| +r}{z_i + |\overrightarrow{z_i} - \overrightarrow{r}| -r},
\end{equation}
\begin{equation}
\theta = \frac{2 \lambda + \alpha_1 + \alpha_2}{2}.
\end{equation}

Durcan found the following $L^2$ orthonormal basis of solutions to $\slashed{D}_x ^\dagger \rho(s) = 0$, where the columns of $\rho$ form the basis: 

\begin{equation}
\rho(s) = \frac{1}{\sqrt{r \sinh (2 r \theta  )}} \left\{
\begin{array}{ll}
e^{(s + \lambda)z_{1}}\sinh{(r \alpha_1)} (z_1  + \slashed{z_1})e^{-\theta \slashed{r}} & \quad s < -\lambda \\
r e^{\frac{2 s + \alpha_1 - \alpha_2}{2} \slashed{r}} & \quad -\lambda < s < \lambda \\
e^{(-s + \lambda)z_{2}}\sinh{(r \alpha_2)} (z_2  - \slashed{z_2})e^{\theta \slashed{r}} & \quad s > \lambda
\end{array}
\right. ,
\end{equation}

\begin{equation}
f_{-\lambda} = \frac{Q^\dagger_{1 -}}{2 |\overrightarrow{r} - \overrightarrow{z_1}|}(\rho(-\lambda_+) - \rho(-\lambda_-)) ,
\end{equation}

\begin{equation}
f_{\lambda} = \frac{Q^\dagger_{2 +}}{2 |\overrightarrow{r} - \overrightarrow{z_2}|}(\rho(\lambda_+) - \rho(\lambda_-)) .
\end{equation}

These satisfy:

\begin{equation}
-\mathbbm{1}\frac{d}{{ds}}\rho(s) - 
(\slashed{T}(s) - \slashed{x}) \rho(s) 
+\delta  (\lambda +s) Q_{1-} f_{-\lambda}
+\delta  (s-\lambda ) Q_{2+} f_{\lambda} = 0 .
\end{equation}

The Nahm equations imply that  ${\slashed{D}^\dagger_x}{\slashed{D}_x}$ is real, thus we can treat the Green's, $G(t,s)$ as a scalar. We find it to be 

\begin{equation}
G(-\lambda \leq t\leq \lambda , s)= \frac {1} {r \text {sinh} (r (2\lambda + \alpha_1 + \alpha_2)} \left(
\begin{array}{cc}
\begin{array}{cc}
e^{z_1 (\lambda +s)} \sinh \left(\alpha _1 r\right)  \sinh \left(r \left(\alpha _2+\lambda -t\right)\right) & s\leq -\lambda  \\
\sinh \left(r \left(\alpha _1+\lambda +s\right)\right) \sinh \left(r \left(\alpha _2+\lambda -t\right)\right) & s>-\lambda \land s\leq t \\
\sinh \left(r \left(\alpha _1+\lambda +t\right)\right) \sinh \left(r \left(\alpha _2+\lambda -s\right)\right) & s>t\land \lambda \geq s \\
e^{z_2 (\lambda -s)}  \sinh \left(r \left(\alpha _1+\lambda +t\right)\right) \sinh \left(\alpha _2 r\right) & s\geq \lambda  \\
\end{array}
\\
\end{array}
\right) ,
\end{equation}

\begin{equation}
G(t<-\lambda , s)= e^{z_1 (\lambda +t)} G(s,-\lambda )	- \frac{1}{z_1} \left(
\begin{array}{cc}
\begin{array}{cc}
e^{(s+\lambda ) z_1} \sinh \left((t+\lambda ) z_1\right) & s\leq t \\
e^{(t+\lambda ) z_1} \sinh \left((s+\lambda ) z_1\right) & s>t\land s\leq -\lambda  \\
\end{array}
\\
\end{array}
\right) ,
\end{equation}

\begin{equation}
G(\lambda <t, s)= e^{z_2 (\lambda -t)} G(s,\lambda )  - \frac{1}{z_2} \left(
\begin{array}{cc}
\begin{array}{cc}
e^{(\lambda -t) z_2} \sinh \left((\lambda -s) z_2\right) & \lambda <s\land s\leq t \\
e^{(\lambda -s) z_2} \sinh \left((\lambda -t) z_2\right) & t\leq s \\
\end{array}
\\
\end{array}
\right)	.
\end{equation}

\section{Solutions to the Dirac equation over  $\mathbb{R}^3$}

Using the solutions to the Weyl equation, the resulting Higgs field $\Phi$ and connection $A_j$ take the form \cite{Durcan}

\begin{equation}
\begin{split}
\Phi &= \frac{1}{2r}(-1 + 2 r \lambda \coth(2 r \theta))\frac{\slashed{r}}{r} 
+ \frac{1}{4 z_2}(1 + \coth(2 r \theta)\frac{\slashed{r}}{r})   
- \frac{1}{4 z_1}(1 - \coth(2 r \theta)\frac{\slashed{r}}{r})   \\
&+ \frac{1}{4 \sinh(2 r \theta)} \frac{\slashed{r}}{r^3}(\frac{\sinh(r \alpha_1)^2}{z_1} [\slashed{z}_1, \slashed{r}] + \frac{\sinh(r \alpha_2)^2}{z_2} [\slashed{z}_2, \slashed{r}]) 
\end{split}
\end{equation}

\begin{equation}
\begin{split}
\overrightarrow{A} &= (\frac{\lambda}{\sinh(2 r \theta)} - \frac{1}{2 r})\frac{[\slashed{r}, d\slashed{x}]}{2 i r}\\
&+ \frac{\sinh(r \alpha_1)}{2 r \sinh(2 r \theta)}( \cosh(r \alpha_1) \frac{[\slashed{r}, d\slashed{x}]}{2 i r} - \sinh(r \alpha_1) \frac{[\slashed{z_1}, d\slashed{x}]}{2 i z_1}) \\	     
&+ \frac{\sinh(r \alpha_2)}{2 r \sinh(2 r \theta)}( \cosh(r \alpha_2) \frac{[\slashed{r}, d\slashed{x}]}{2 i r} - \sinh(r \alpha_2) \frac{[\slashed{z_2}, d\slashed{x}]}{2 i z_2}) \\
&+ \frac{(\overrightarrow{r} \times \overrightarrow{z_1})}{r z_1 }\frac{\sinh(r \alpha_1)^2}{2 r \cosh(r \theta)} e^{-\theta \slashed{r}} \\
&- \frac{(\overrightarrow{r} \times \overrightarrow{z_2})}{r z_2 }\frac{\sinh(r \alpha_2)^2}{2 r \cosh(r \theta)} e^{\theta \slashed{r}}
\end{split}
\end{equation}

In order to write out the solution to the Dirac equation compactly, we build it up from a few pieces. Viewing $G_x(t,s)$ as a scalar, the solution to the Dirac equation is given\cite{Schenk} by

\begin{equation}
\begin{split}
\chi_t = \begin{pmatrix}
\int \rho_x(s)^{\dagger}  G_x(t,s)  \, ds & 0 \\
0 & \int \rho_x(s)^{\dagger}  G_x(t,s)  \, ds
\end{pmatrix}
\begin{pmatrix}
0 \\
1 \\
-1 \\
0 \\
\end{pmatrix}
\end{split}
\end{equation}

This automatically satisfies the Dirac equation

\begin{equation}
D_t \chi_t = 0 ,
\end{equation}

and its $L^2$ norm is
\begin{equation}
\int \chi^\dagger_t \chi_t \, dx = \frac{\pi}{2}. 
\end{equation}

Here $\int \rho_x(s)^{\dagger}  G_x(t,s)  \, ds$ is a 2x2 matrix. To write it explicitly, we introduce the following 

\begin{equation} \label{crazyPsi}
\begin{split}
\Psi(t) &= \frac{1}{2 \sqrt{r \sinh (2 r (\lambda + \frac{\alpha_1 + \alpha_2}{2})  )}}( \\
&(-t \sinh(r (t + \frac{\alpha_1 - \alpha_2}{2})) + \lambda \cosh(r (t + \frac{\alpha_1 - \alpha_2}{2})) \tanh(r (\lambda + \frac{\alpha_1 + \alpha_2}{2}))) \\
+&(-t \cosh(r (t + \frac{\alpha_1 - \alpha_2}{2})) + \lambda \sinh(r (t + \frac{\alpha_1 - \alpha_2}{2})) \coth(r (\lambda + \frac{\alpha_1 + \alpha_2}{2})))\frac{\slashed{r}}{r}  \\
&- \frac{\sinh(r \alpha_1) \sinh(r (-t + \lambda + \alpha_2))}{r \sinh(r (\lambda + \frac{\alpha_1 + \alpha_2}{2}))}e^{-(\lambda + \frac{\alpha_1 + \alpha_2}{2}) \slashed{r}}( \cosh(r \alpha_1) \frac{\slashed{r}}{r} - \sinh(r \alpha_1) \frac{\slashed{z_1}}{z_1})   \\
&+ \frac{\sinh(r \alpha_2) \sinh(r (+t + \lambda + \alpha_1))}{r \sinh(r (\lambda + \frac{\alpha_1 + \alpha_2}{2}))}e^{+(\lambda + \frac{\alpha_1 + \alpha_2}{2}) \slashed{r}}( \cosh(r \alpha_2) \frac{\slashed{r}}{r} - \sinh(r \alpha_2) \frac{\slashed{z_2}}{z_2})) \\
\end{split}
.
\end{equation}

Then $\int \rho_x(s)^{\dagger}  G_x(t,s)  \, ds$ can be written as

\begin{equation}
\begin{split}
\int \rho_x(s)^{\dagger}  G_x(t,s)  \, ds =  \left\{
\begin{array}{ll}
e^{(+t + \lambda)z_{1}}\Psi(-\lambda) - \frac{(+t+\lambda)}{2 z_1}\rho(t)^{\dagger} & \quad t < -\lambda \\
\Psi(t) & \quad -\lambda < t < \lambda \\
e^{(-t + \lambda)z_{2}}\Psi(+\lambda) - \frac{(-t+\lambda)}{2 z_2}\rho(t)^{\dagger} & \quad t > \lambda
\end{array}
\right.
\end{split}.
\end{equation}

Note that while the solutions to Weyl equation are discontinuous across $\lambda$, the solutions to Dirac equation change continiously as we vary $t$ across $\lambda$. Additionally, each piece of the solution remains in the kernel of the Dirac operator even across $\lambda$, though they fail to be $L^2$ over $\mathbb{R}^3$. Combining all of this, we can write solution to the Dirac equation as: 

\begin{equation}
\begin{split}
\chi_t = \begin{pmatrix}
\left\{
\begin{array}{ll}
e^{(t + \lambda)z_{1}}\Psi(-\lambda) - \frac{(t+\lambda)}{2 z_1}\rho(t)^{\dagger} & \quad t < -\lambda \\
\Psi(t) & \quad -\lambda < t < \lambda \\
e^{(-t + \lambda)z_{2}}\Psi(\lambda) - \frac{(-t+\lambda)}{2 z_2}\rho(t)^{\dagger} & \quad t > \lambda
\end{array}
\right. & 0 \\
0 & \left\{
\begin{array}{ll}
e^{(t + \lambda)z_{1}}\Psi(-\lambda) - \frac{(t+\lambda)}{2 z_1}\rho(t)^{\dagger} & \quad t < -\lambda \\
\Psi(t) & \quad -\lambda < t < \lambda \\
e^{(-t + \lambda)z_{2}}\Psi(\lambda) - \frac{(-t+\lambda)}{2 z_2}\rho(t)^{\dagger} & \quad t > \lambda
\end{array}
\right.
\end{pmatrix}
\begin{pmatrix}
0 \\
1 \\
-1 \\
0 \\
\end{pmatrix}
\end{split}
\end{equation}

Introducing the following operation

\begin{equation}
\begin{split}
\begin{pmatrix}
\left.
\begin{array}{ll}
a & b\\
c & d 
\end{array}
\right.
\end{pmatrix}^C
=
\begin{pmatrix}
b \\
d \\
-a \\
-c \\
\end{pmatrix}
\end{split},
\end{equation}

We can write the solution to Dirac equation as:

\begin{equation}
\begin{split}
\chi_t = 
\left\{
\begin{array}{ll}
(e^{(+t + \lambda)z_{1}}\Psi(-\lambda) - \frac{(+t+\lambda)}{2 z_1}\rho(t)^{\dagger})^C & \quad t < -\lambda \\
\Psi(t)^C & \quad -\lambda < t < \lambda \\
(e^{(-t + \lambda)z_{2}}\Psi(+\lambda) - \frac{(-t+\lambda)}{2 z_2}\rho(t)^{\dagger})^C & \quad t > \lambda
\end{array}
\right.
\end{split}. 
\end{equation}

Here we restate $\rho(t)$ for convenience

\begin{equation}
\rho(t) = \frac{1}{\sqrt{r \sinh ( r (2\lambda + \alpha_1 + \alpha_2)  )}} \left\{
\begin{array}{ll}
e^{(+t + \lambda)z_{1}}\sinh{(r \alpha_1)} (z_1  + \slashed{z_1})e^{-(\lambda + \frac{\alpha_1 + \alpha_2}{2}) \slashed{r}} & \quad t < -\lambda \\
e^{(-t + \lambda)z_{2}}\sinh{(r \alpha_2)} (z_2  - \slashed{z_2})e^{+(\lambda + \frac{\alpha_1 + \alpha_2}{2}) \slashed{r}} & \quad t > \lambda
\end{array}
\right. .
\end{equation}

\section{Observations}

Earlier it was noted that this configuration can be viewed as a combination of two Dirac singularities and one BPS monopole. In the following, we remove singularities/monopoles and observe the resulting solution to the Dirac equation. The author suspects that the results of Section \ref{bps} and Section \ref{bps1sing} are already known, but was unable to find them in the literature. 

\subsection{Dirac Monopole}

We take the limit $\overrightarrow{T}_{tH}, {T}_{D_1} \rightarrow \infty$. The solution to Dirac equation becomes

\begin{equation}
\begin{split}
\chi_t = \frac{(t - \lambda ) e^{ (\lambda - t)z_2}}{4 \sqrt{z_2 - \overrightarrow{z_2}\cdot \hat{r} }} \begin{pmatrix}
1 + \frac{\slashed{r}}{r} & 0 \\
0 & 1 + \frac{\slashed{r}}{r} \\
\end{pmatrix}
\begin{pmatrix}
1 - \frac{\slashed{z_2}}{z_2} & 0 \\
0 & 1 - \frac{\slashed{z_2}}{z_2} \\
\end{pmatrix}
\begin{pmatrix}
0 \\
1 \\
-1 \\
0 \\
\end{pmatrix}
\end{split}
\end{equation}

Under gauge transformation this becomes

\begin{equation}\label{diracmonopoleSolution}
\begin{split}
\chi_t = \frac{(t - \lambda ) e^{ (\lambda - t)z_2}}{2 z_2 \sqrt{z_2 - z_2^3 }} 
\begin{pmatrix}
-z_2^1 + iz_2^2 \\
0 \\
-z_2 + z_2^3 \\
0 \\
\end{pmatrix}
\end{split}
\end{equation}

And the Higgs field takes the form

\begin{equation}
\Phi = \begin{pmatrix}
\lambda + \frac{1}{2 z_2} & 0 \\
0 & -\lambda \\
\end{pmatrix}
\end{equation}

Inspired by \cite{Cheng}, we convert our solution to spherical coordinates, $z_2^1 = z_2 \sin{\theta}\sin{\phi}, z_2^2 = z_2 \sin{\theta}\cos{\phi}, z_2^3 = z_2 \cos{\theta}$. Then the solution to Dirac equation becomes:

\begin{equation}\label{diracmonopoleSolutionSpherical}
\begin{split}
\chi_t = \frac{(t - \lambda ) e^{ (\lambda - t)z_2}  }{\sqrt{2 z_2} } 
\begin{pmatrix}
i e^{i \phi} \cos{\frac{\theta}{2}}\\
0 \\
-\sin{\frac{\theta}{2}} \\
0 \\
\end{pmatrix}
\end{split}
\end{equation}

\subsubsection{Comparison with Cheng and Ford}

In \cite{Cheng}, Cheng and Ford computed explicit solutions to Dirac equation in the abelian case. We recover their solution in the case of one singularity by taking the limit  $\overrightarrow{T}_{tH}, {T}_{D_2} \rightarrow \infty$. The Higgs field takes the form 

\begin{equation}
\Phi = \begin{pmatrix}
\lambda  & 0 \\
0 & -\lambda -  \frac{1}{2 z_1} \\
\end{pmatrix}
\end{equation}

The solution to Dirac equation in spherical coordinates becomes

\begin{equation}\label{diracmonopoleFordSolutionSpherical}
\begin{split}
\chi_t = \frac{-i (t + \lambda )}{\sqrt{2}} \frac{ e^{ (t + \lambda)z_1}  }{\sqrt{ z_1} } 
\begin{pmatrix}
0 \\
-\sin{\frac{\theta}{2}} \\
0 \\
e^{i \phi} \cos{\frac{\theta}{2}} \\
\end{pmatrix}
\end{split}
\end{equation}

Which matches up to over all normalization the results of Cheng and Ford: Equation (5) of \cite{Cheng} 

\subsubsection{Comparison with direct $U(1)$ solution}

Alternatively, if we had started with the following Nahm data

\begin{equation}
\overrightarrow{T}(s) =  \left\{
\begin{array}{ll}
\overrightarrow{T}_{D_2}  \in \mathbb{R}^3 & \quad s > \lambda
\end{array}
\right.
\end{equation}

Note that $\overrightarrow{T}(s) $ is only defined for $s > \lambda$. We could find solutions to the Weyl operator and perform the Nahm transform. Then the Higgs field would be

\begin{equation}
\Phi = 
\lambda + \frac{1}{2 z_2}
,
\end{equation}

and the solution to the resulting Dirac equation is given by

\begin{equation}\label{monopoleSpinorReduced}
\begin{split}
\chi_t = \frac{(t - \lambda ) e^{ (\lambda - t)z_2}}{2 z_2 \sqrt{z_2 - z_2^3 }} 
\begin{pmatrix}
-z_2^1 + iz_2^2 \\
-z_2 + z_2^3 \\
\end{pmatrix}
\end{split},
\end{equation}

which matches the result of \eqref{diracmonopoleSolution}. 

\subsection{BPS monopole} \label{bps}

The symbol $\alpha_i$ allows us to easily compute the limits for moving $ {T}_{D_i} \rightarrow \infty$ by sending  $\alpha_i \rightarrow 0$. $\Psi(t)$ in \eqref{crazyPsi} contains terms proportional to $\sinh(r \alpha_i$) in the last two lines which vanish when sending  $\alpha_1, \alpha_2 \rightarrow 0$ respectively. \\
We can recover the solution to Dirac equation for an $SU(2)$ monopole. At $t=0$, $\Psi$ takes the form

\begin{equation}
\Psi(0) = \frac{\lambda \tanh (r \lambda )}{2 \sqrt{r \sinh (2 r \lambda  )}}. 
\end{equation}

Which leads to

\begin{equation}\label{BPSsolution}
\chi_0 = \frac{\lambda \tanh (r \lambda )}{2 \sqrt{r \sinh (2 r \lambda  )}}
\begin{pmatrix}
0 \\
1 \\
-1 \\
0 \\
\end{pmatrix}
.
\end{equation}

The Higgs field and Connection take the following form: 

\begin{equation}
\begin{split}
\Phi &= \frac{1}{2r}(-1 + 2 r \lambda \coth(2 r \lambda))\frac{\slashed{r}}{r} 
\end{split}
,
\end{equation}

\begin{equation}
\begin{split}
\overrightarrow{A} &= (\frac{\lambda}{\sinh(2 r \lambda)} - \frac{1}{2 r})\frac{[\slashed{r}, d\slashed{x}]}{2 i r}
\end{split}
.
\end{equation}

\subsection{BPS monopole and 1 Coincident Singularity} \label{bps1sing}

We can recover the solution to Dirac equation for an $SU(2)$ monopole with one singularity by setting $T_{D_1}  = T_{D_2}$. At $t=0$ our solution takes the form

\begin{equation}
\chi_0 = \frac{\tanh (r (\lambda + \alpha))}{2 \sqrt{r \sinh (2 r (\lambda + \alpha)  )}}(\lambda + \frac{1}{2z} + \frac{\sinh(r \alpha)^2}{2 z r^2}[\slashed{z}, \slashed{r}])
\begin{pmatrix}
0 \\
1 \\
-1 \\
0 \\
\end{pmatrix}
.
\end{equation}

With Higgs field given by:

\begin{equation}
\begin{split}
\Phi &= \frac{1}{2r}(-1 + 2 r \lambda \coth(2 r \theta))\frac{\slashed{r}}{r} 
+ \frac{1}{2 z}( \coth(2 r \theta)\frac{\slashed{r}}{r})  
+ \frac{1}{2 \sinh(2 r \theta)} \frac{\slashed{r}}{r^3}
(\frac{\sinh(r \alpha)^2}{z} [\slashed{z}, \slashed{r}])
\end{split}
.
\end{equation}

\subsection{Effect of twisting/moving the monopoles on point wise norm}

The following graphs represent the pointwise norm, $\chi_t ^{\dagger} \chi_t$, of the solution to Dirac equation. The monopole and singularities are arranged in a line and we observe the norm on a plane, parallel to the line, above them. The blue lines passing perpindicularly through the plane passes through the location of the monopole and singularities. The negative singularity is on the left, the BPS monopole is in the center and the positive singularity is on the right. The Dirac operator 	$D_t = -(\Phi -t) +   \overset{\rightharpoonup }{\sigma } \otimes (\frac{d}{d \overset{\rightharpoonup }{x}} - i \overset{\rightharpoonup }{A})$ has a parameter $t$ which we vary in the second set of graphs.

\newpage

\subsubsection{Moving the Positive Dirac Singularity Plots}

\begin{figure}[h!]
	\centering
	\begin{subfigure}[b]{0.4\linewidth}
		\includegraphics[width=\linewidth]{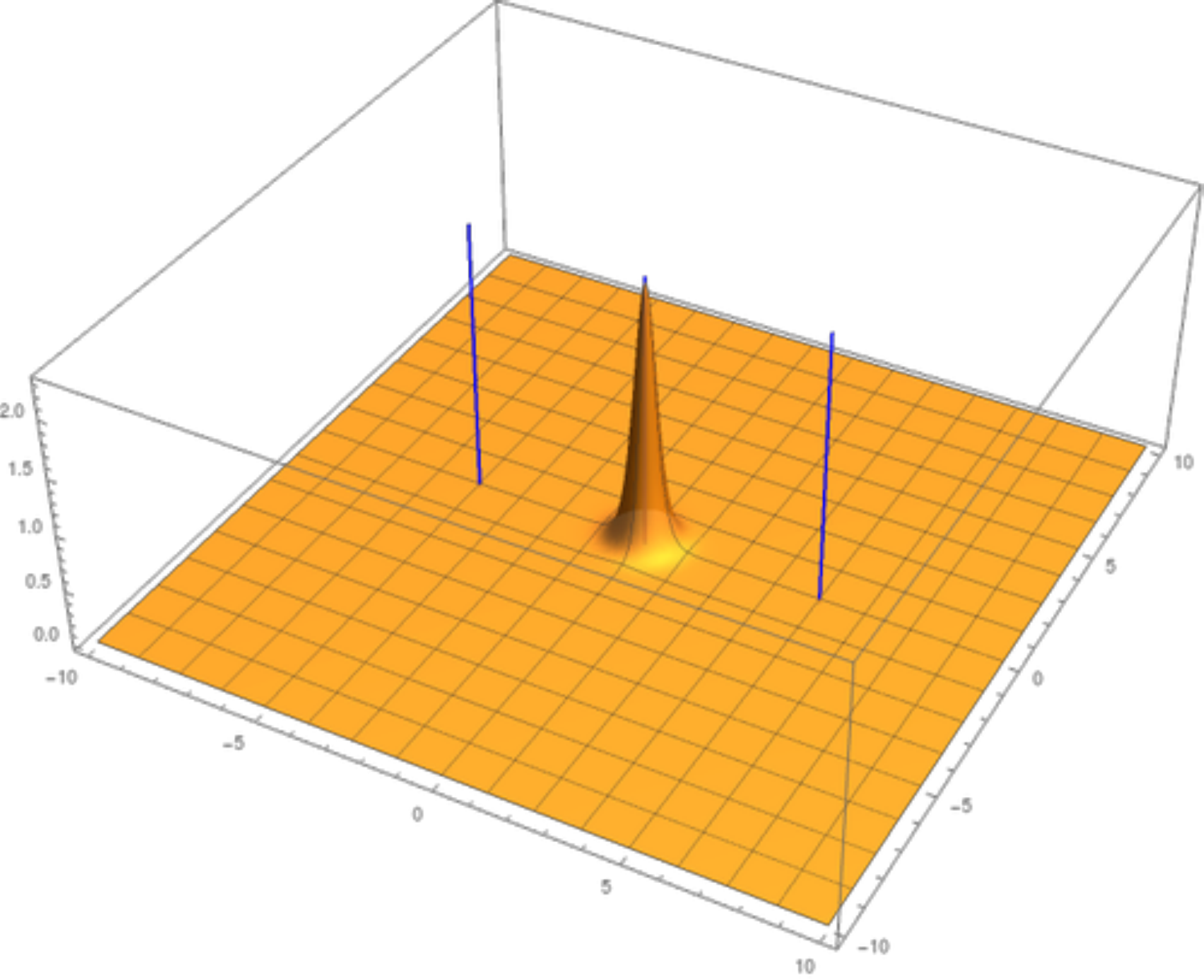}
		\caption{\label{fig:i} $\Delta = 5$}
	\end{subfigure}
	\begin{subfigure}[b]{0.4\linewidth}
		\includegraphics[width=\linewidth]{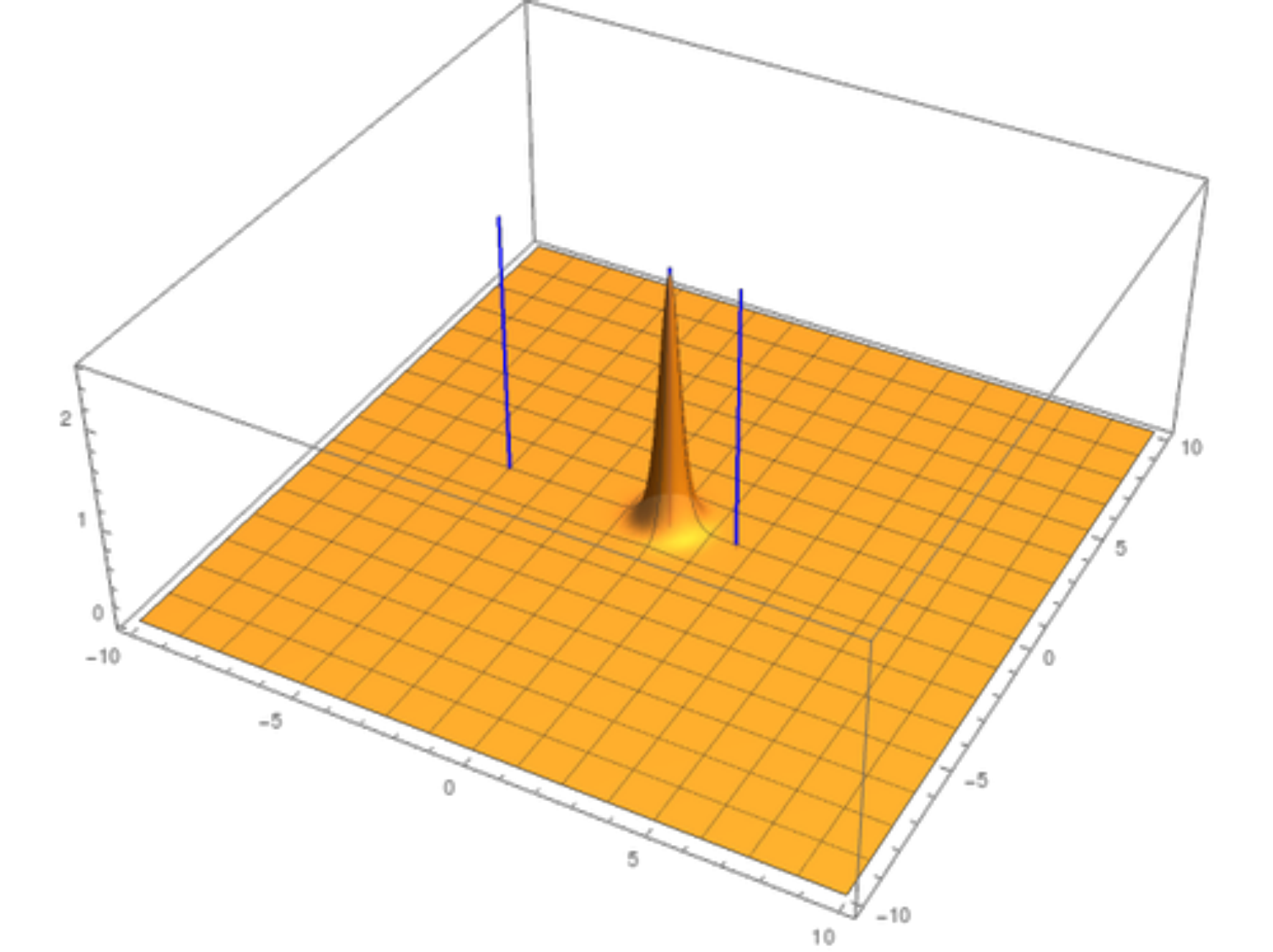}
		\caption{\label{fig:i}$\Delta = 2$}
	\end{subfigure}
	\begin{subfigure}[b]{0.4\linewidth}
		\includegraphics[width=\linewidth]{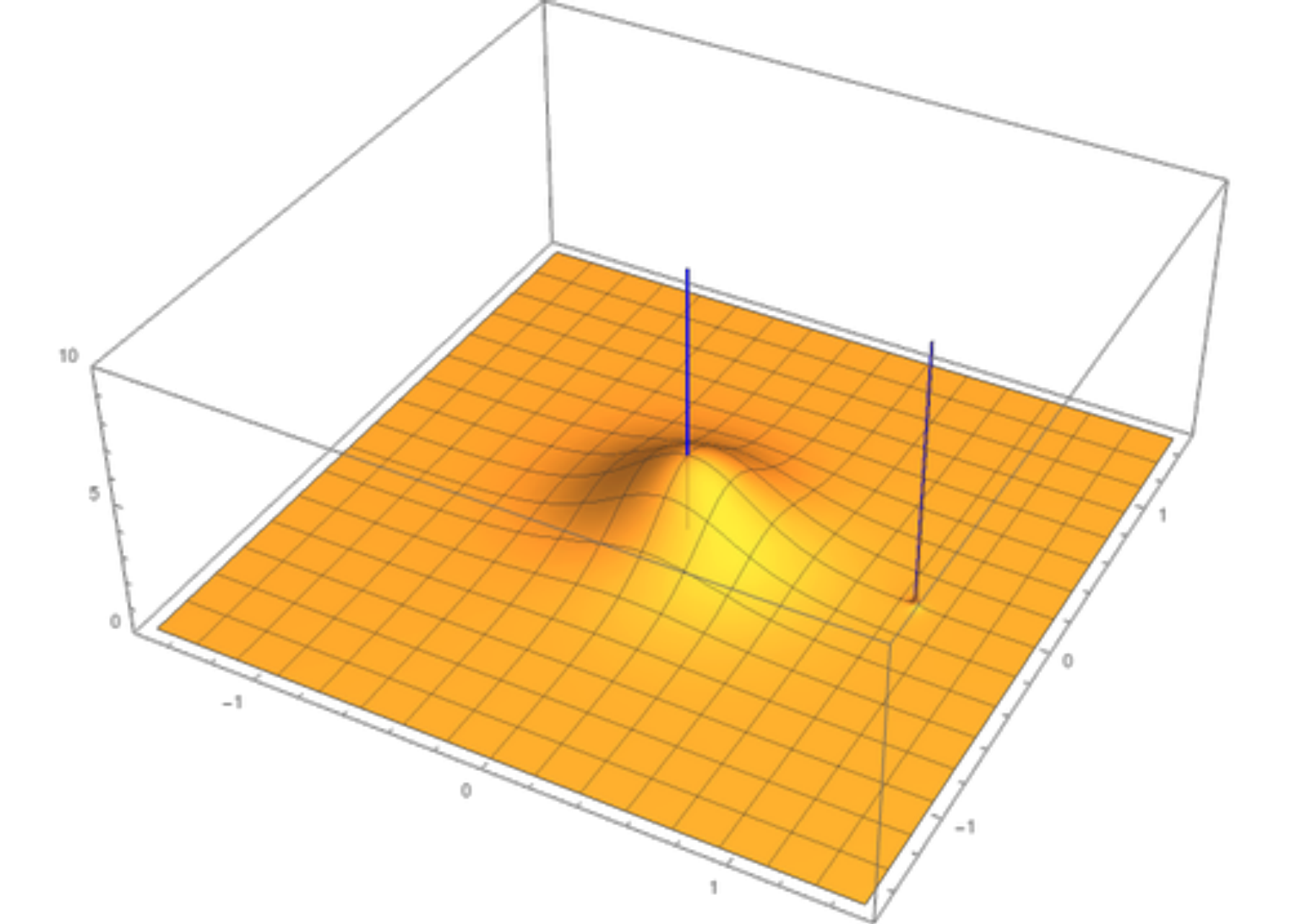}
		\caption{\label{fig:i} Zoomd out, $\Delta = 1$}
	\end{subfigure}
	\begin{subfigure}[b]{0.4\linewidth}
		\includegraphics[width=\linewidth]{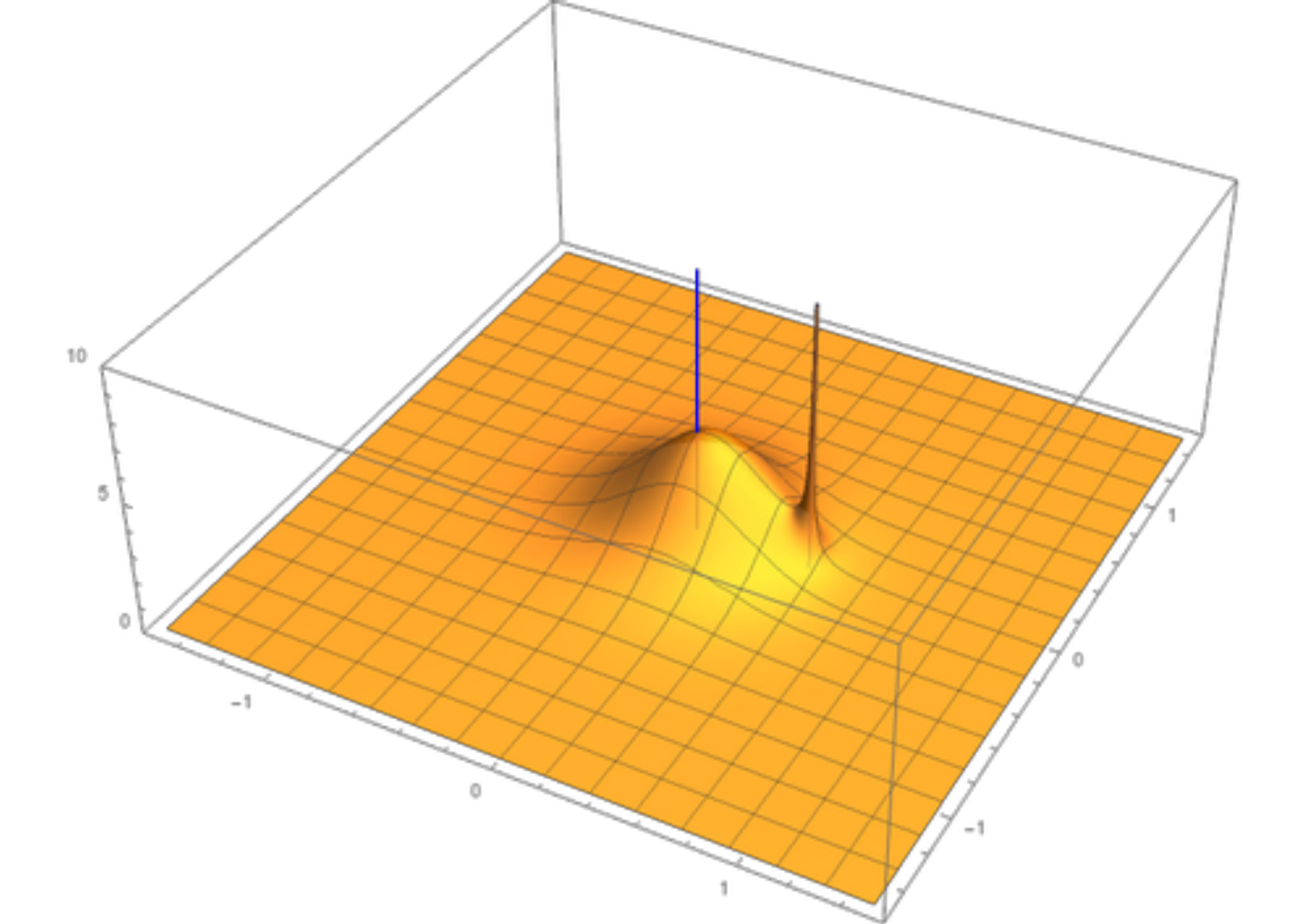}
		\caption{\label{fig:i}  $\Delta = 0.5$}
	\end{subfigure}
	
	\begin{subfigure}[b]{0.4\linewidth}
		\includegraphics[width=\linewidth]{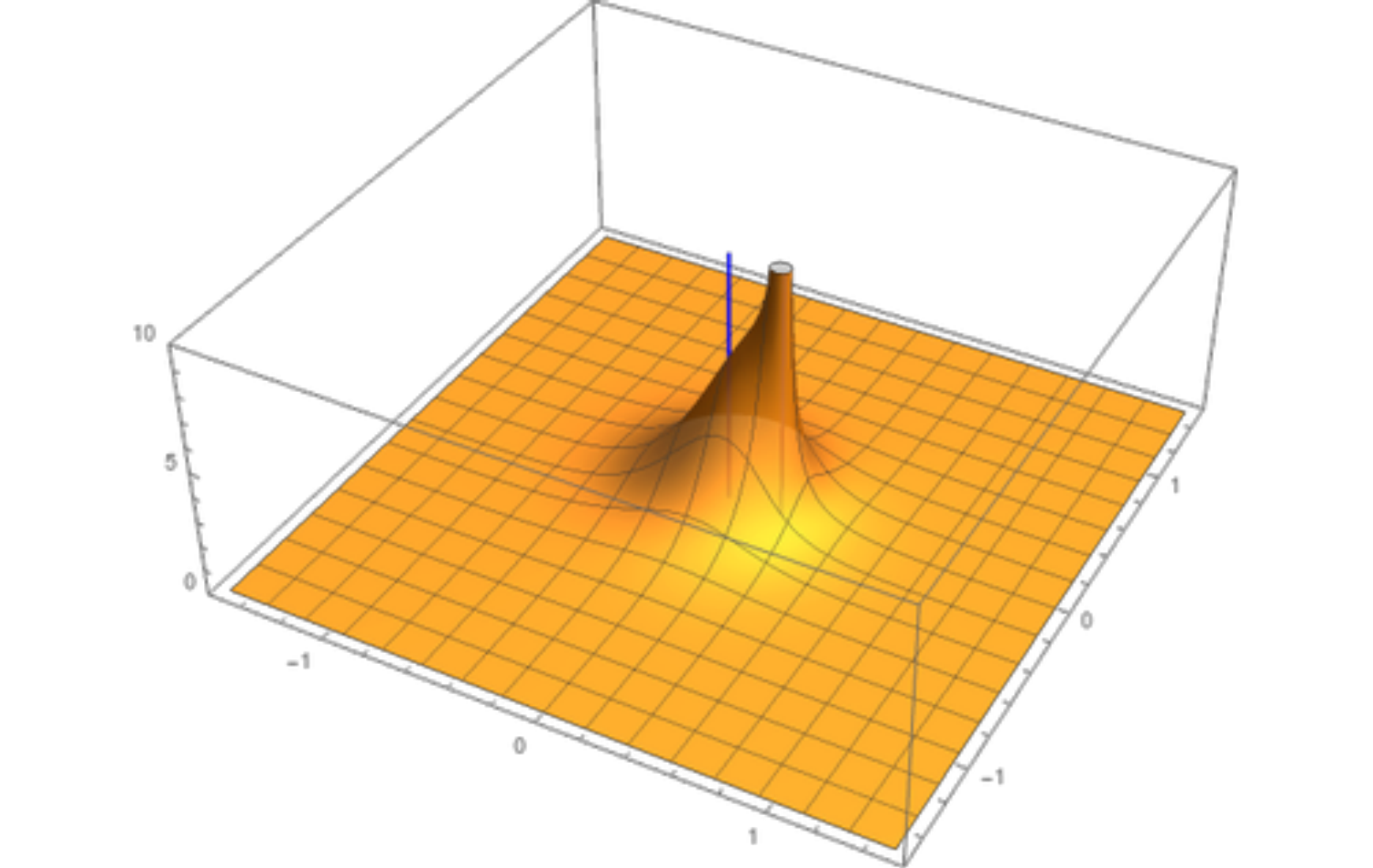}
		\caption{\label{fig:i}  $\Delta = 0.25$}
	\end{subfigure}

	\caption{Norm of the Dirac solutions while moving the positive singularity towards the BPS monopole}
	\label{fig:Movment}
\end{figure}

In Figure \ref{fig:Movment}, we observe how the pointwise norm changes as we move the positive singularity towards the BPS monopole. When far away, there is a distinct region corresponding to the location of the  BPS monopole. As we move the positive singularity towards the BPS monopole, we can still spot two distinct regions corresponding to the singularity and the BPS monopole, but as they get very close, the singularity seems to spread out and the distinct regions seem to merge. In these plots, $t = 0, \lambda = 2$. The symbol $\Delta$ will denote distance between the positive singularity and the BPS monopole.

\newpage

\subsubsection{Twisted Plots}

\begin{figure}[h!]
	\centering
	\begin{subfigure}[b]{0.3\linewidth}
	\includegraphics[width=\linewidth]{NoTwistFarAway.pdf}
\caption{\label{fig:i} $t=0$}
	\end{subfigure}
	\begin{subfigure}[b]{0.3\linewidth}
	\includegraphics[width=\linewidth]{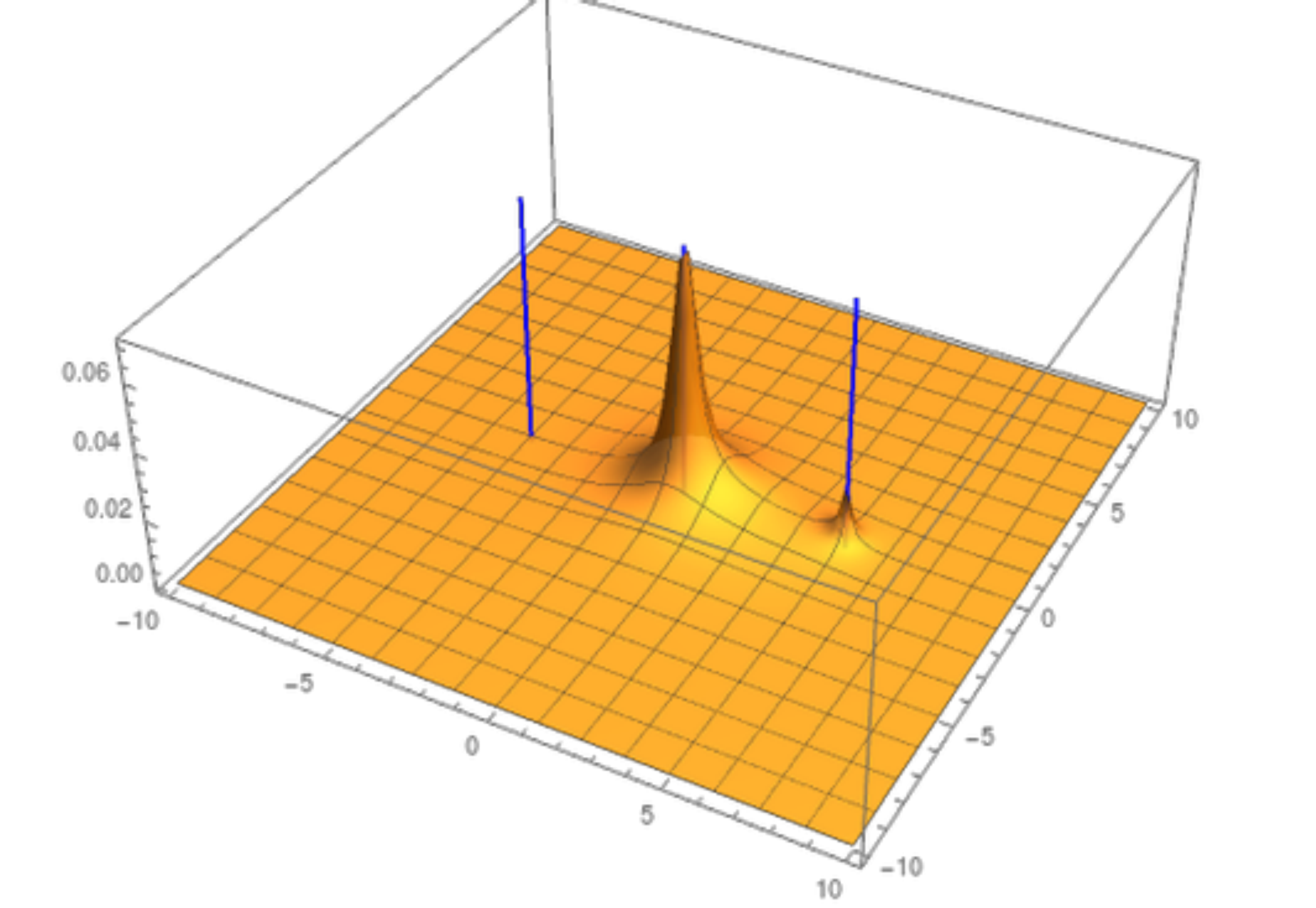}
\caption{\label{fig:i} $t=1$}
	\end{subfigure}
	\begin{subfigure}[b]{0.3\linewidth}
	\includegraphics[width=\linewidth]{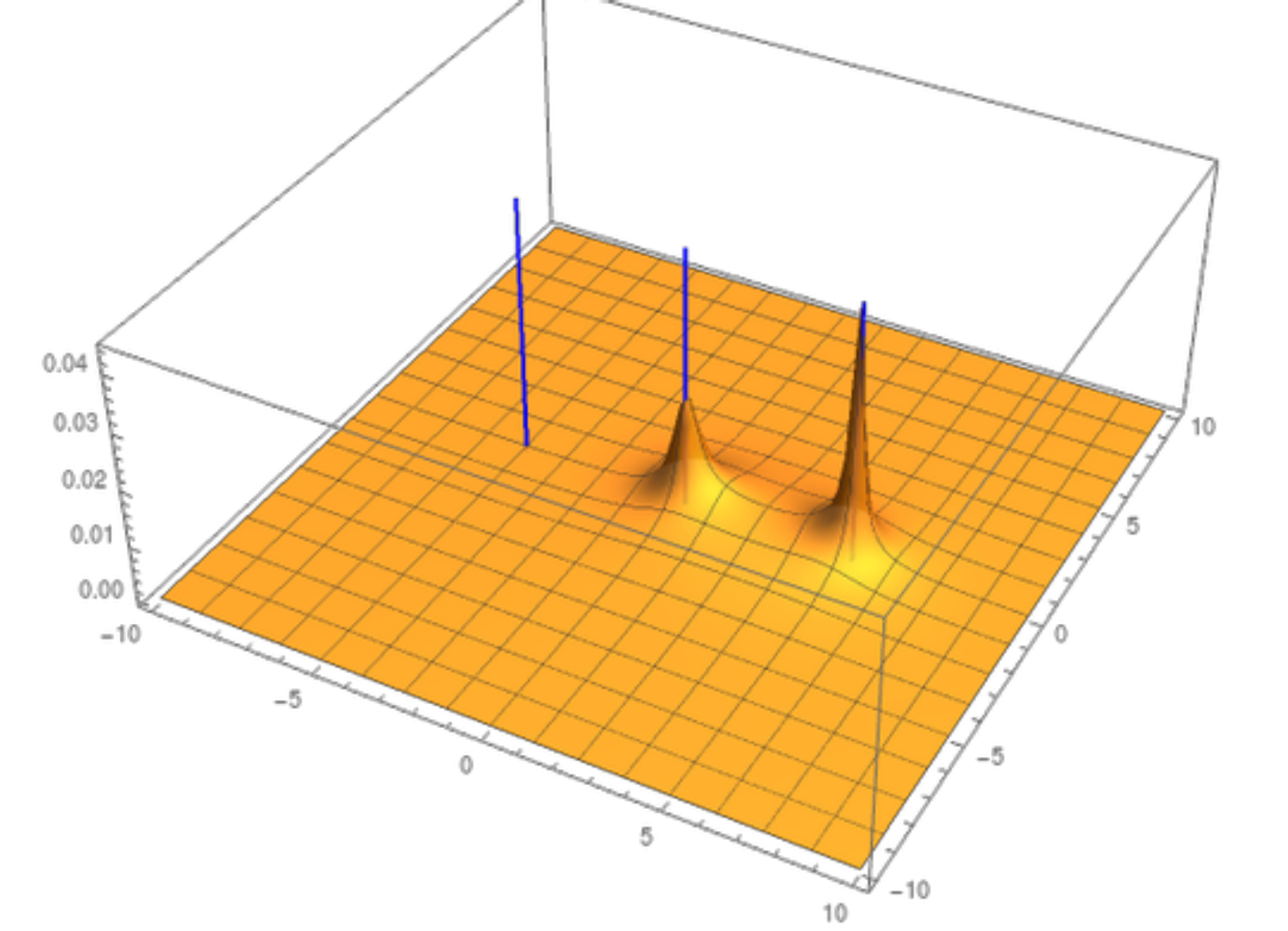}
\caption{\label{fig:i} $t=1.99999$}
	\end{subfigure}
	\begin{subfigure}[b]{0.3\linewidth}
	\includegraphics[width=\linewidth]{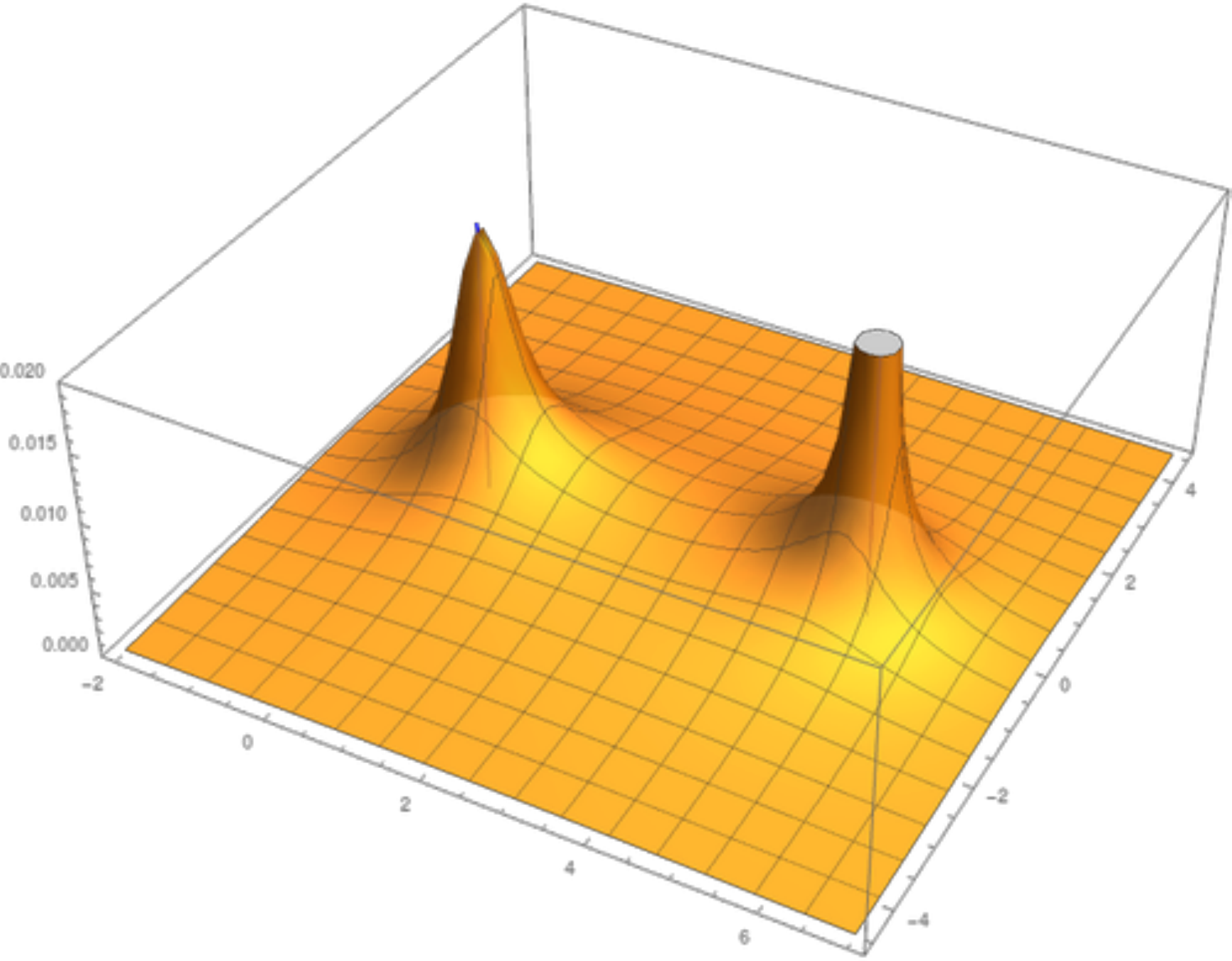}
\caption{\label{fig:i} Zoom in of previous case, $t=1.99999$}
	\end{subfigure}
	
	\begin{subfigure}[b]{0.3\linewidth}
	\includegraphics[width=\linewidth]{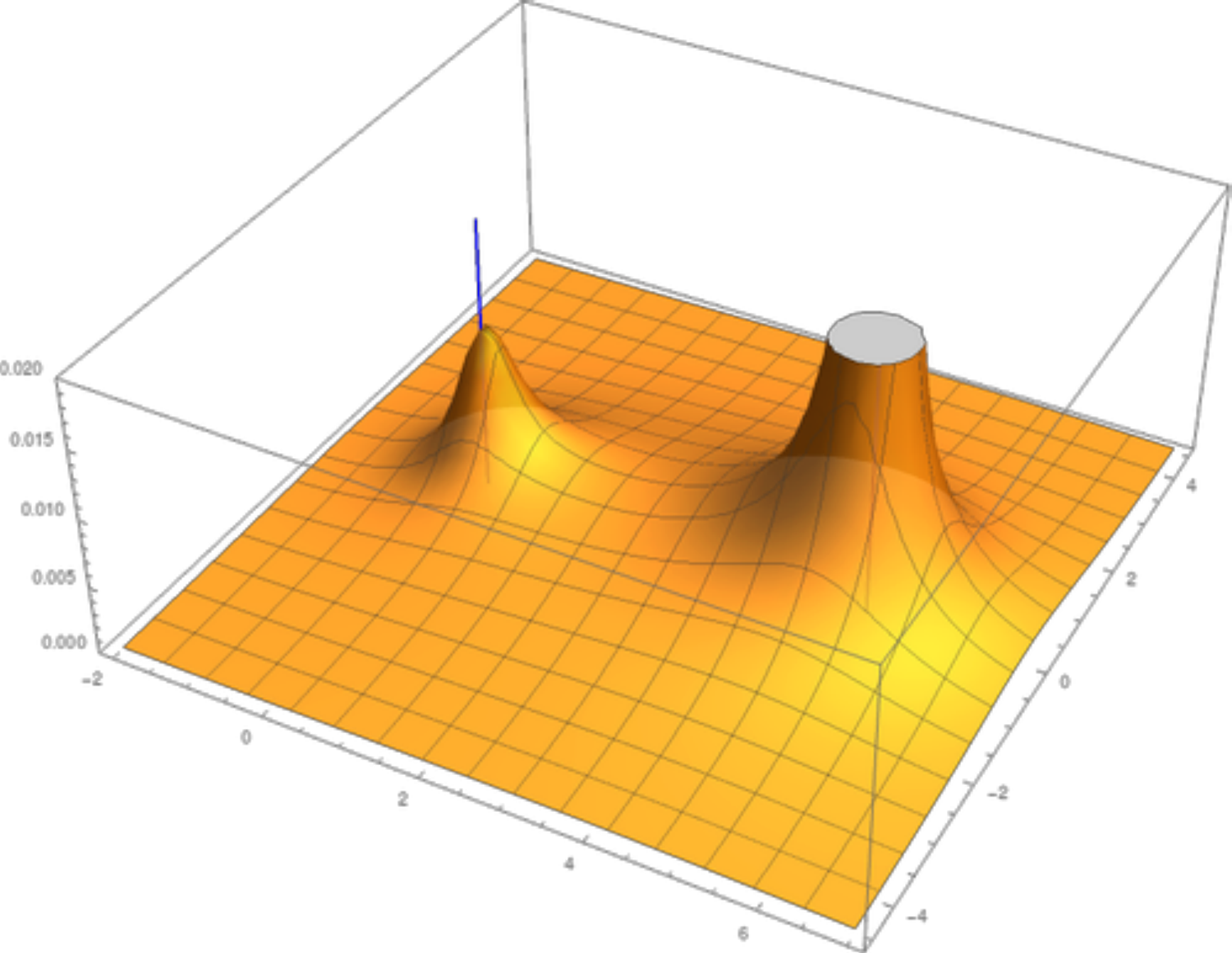}
\caption{\label{fig:i}  $t=2.05$}
	\end{subfigure}
		\begin{subfigure}[b]{0.3\linewidth}
	\includegraphics[width=\linewidth]{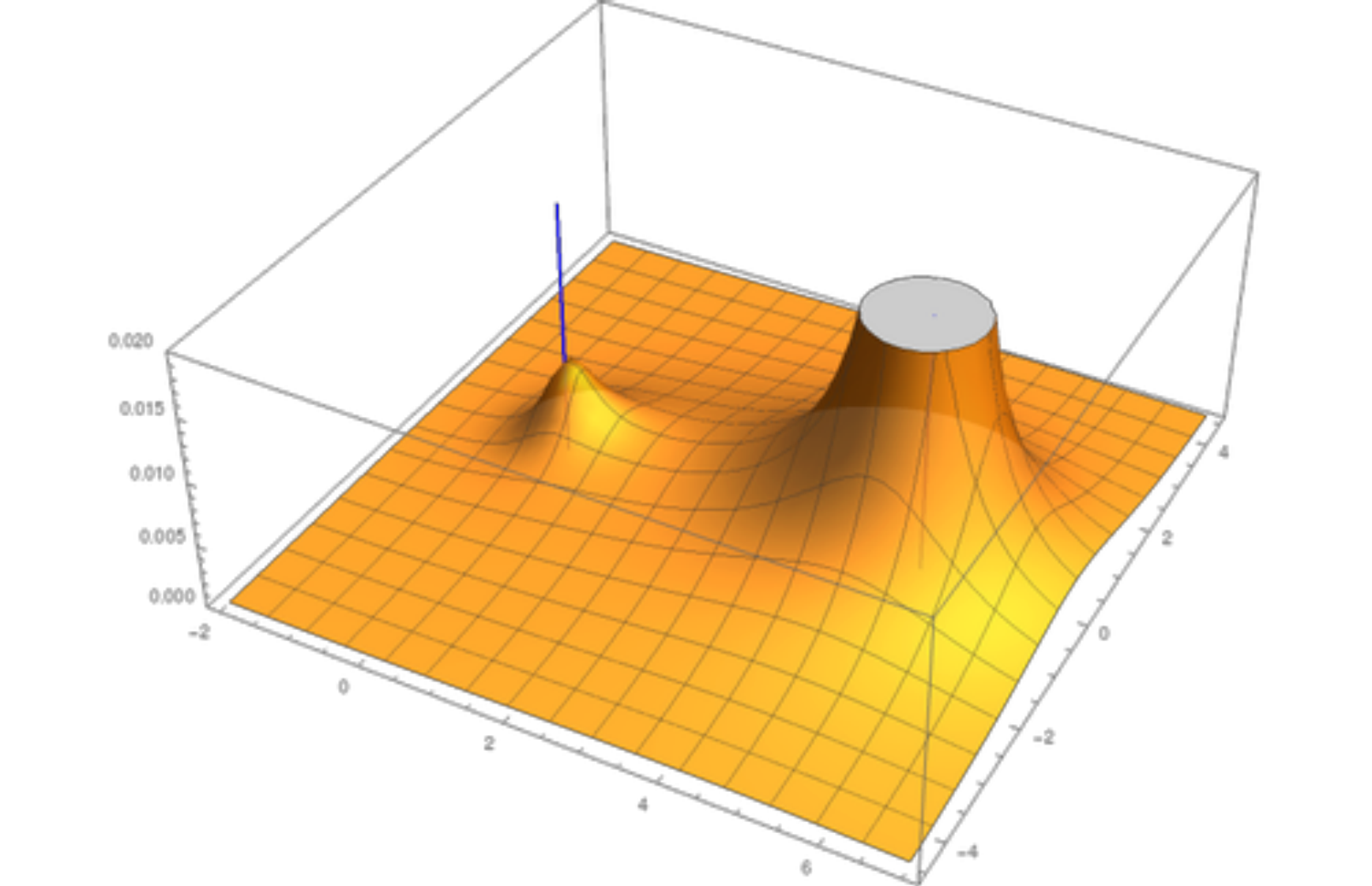}
\caption{\label{fig:i}  $t=2.1$}
	\end{subfigure}
		\begin{subfigure}[b]{0.3\linewidth}
	\includegraphics[width=\linewidth]{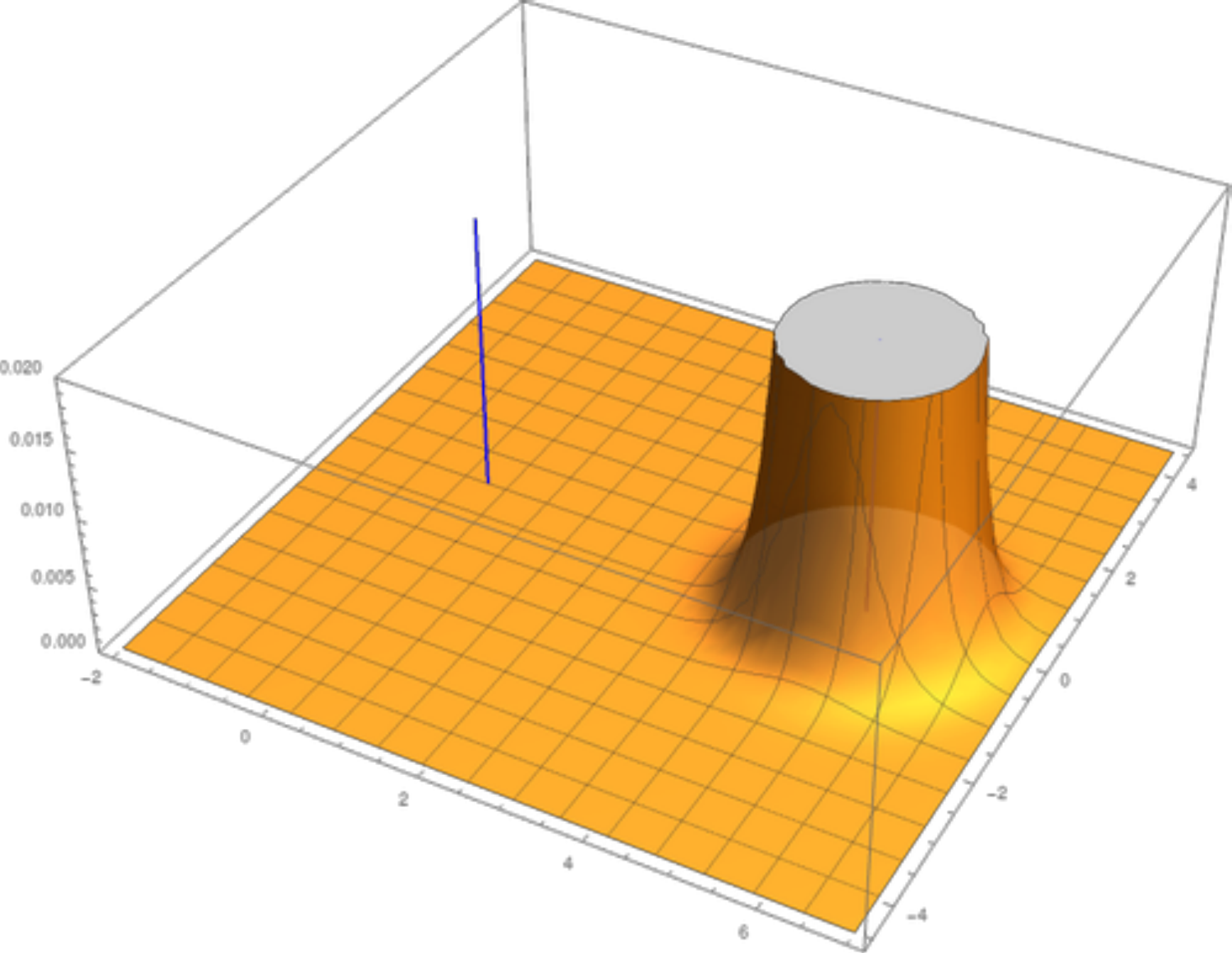}
\caption{\label{fig:i} $t=4$}
\end{subfigure}
		\begin{subfigure}[b]{0.3\linewidth}
	\includegraphics[width=\linewidth]{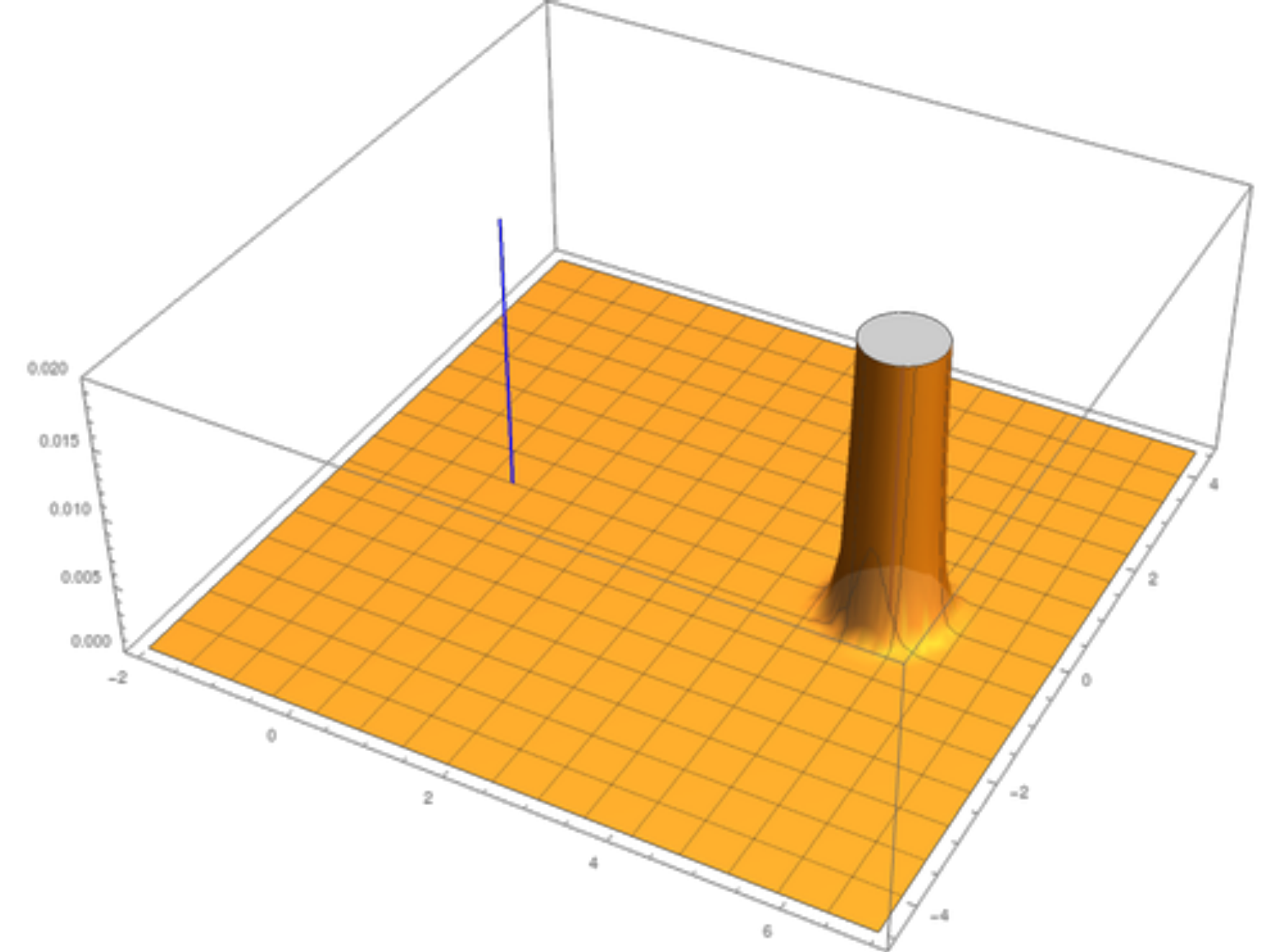}
\caption{\label{fig:i} $t=10$}
\end{subfigure}

	\caption{Twisting the solution to Dirac Equation.}
	\label{fig:Twisting}
\end{figure}

In Figure \ref{fig:Twisting}, we twist the solution by modifying the value of $t$. Specifically, we increase the value of $t$ with the initial plot corresponding to $t=0$. As $t$ increased, the pointwise norm near the BPS monopole decreases. Near the positive singularity, the region in which the pointwise norm is "large" initially increases, and then the norm seems to concentrate at the location of the singularity. In these plots, we set $\lambda = 2$.

\newpage

\acknowledgments

The author would like to thank Sergey Cherkis for his guidance during this project. 

\newpage 

\section{Appendix}

This page contains, in one place, all necessary formula and symbols. 

$\overrightarrow{T}_{D_i}$ location of Dirac singularity. $\overrightarrow{T}_{'tHP}$ location of BPS monopole. 

\begin{equation}
\overrightarrow{z_i} =\overrightarrow{x} -\overrightarrow{T}_{D_i} ,
\end{equation}
\begin{equation}
\overrightarrow{r} =\overrightarrow{x} - \overrightarrow{T}_{'tHP} ,
\end{equation}

\begin{equation}
e^{2 r \alpha _i}=\frac{z_i + |\overrightarrow{z_i} - \overrightarrow{r}| +r}{z_i + |\overrightarrow{z_i} - \overrightarrow{r}| -r} .
\end{equation}

\begin{equation}
\begin{split}
\Psi(t) &= \frac{1}{2 \sqrt{r \sinh (2 r (\lambda + \frac{\alpha_1 + \alpha_2}{2})  )}}( \\
& \quad \; (-t \sinh(r (t + \frac{\alpha_1 - \alpha_2}{2})) + \lambda \cosh(r (t + \frac{\alpha_1 - \alpha_2}{2})) \tanh(r (\lambda + \frac{\alpha_1 + \alpha_2}{2}))) \\
&+ (-t \cosh(r (t + \frac{\alpha_1 - \alpha_2}{2})) + \lambda \sinh(r (t + \frac{\alpha_1 - \alpha_2}{2})) \coth(r (\lambda + \frac{\alpha_1 + \alpha_2}{2})))\frac{\slashed{r}}{r}  \\
&- \frac{\sinh(r \alpha_1) \sinh(r (\lambda -t + \alpha_2))}{r \sinh(r (\lambda + \frac{\alpha_1 + \alpha_2}{2}))}e^{-(\lambda + \frac{\alpha_1 + \alpha_2}{2}) \slashed{r}}( \cosh(r \alpha_1) \frac{\slashed{r}}{r} - \sinh(r \alpha_1) \frac{\slashed{z_1}}{z_1})   \\
&+ \frac{\sinh(r \alpha_2) \sinh(r (\lambda + t + \alpha_1))}{r \sinh(r (\lambda + \frac{\alpha_1 + \alpha_2}{2}))}e^{+(\lambda + \frac{\alpha_1 + \alpha_2}{2}) \slashed{r}}( \cosh(r \alpha_2) \frac{\slashed{r}}{r} - \sinh(r \alpha_2) \frac{\slashed{z_2}}{z_2})) , \\ 
\end{split}
\end{equation}

\begin{equation}
\rho(t)^{\dagger} = \frac{1}{\sqrt{r \sinh (2 r (\lambda + \frac{\alpha_1 + \alpha_2}{2})  )}} \left\{
\begin{array}{ll}
e^{(\lambda + t)z_{1}}\sinh{(r \alpha_1)}e^{-(\lambda + \frac{\alpha_1 + \alpha_2}{2}) \slashed{r}} (z_1  + \slashed{z_1}) & \quad t < -\lambda \\
e^{( \lambda - t)z_{2}}\sinh{(r \alpha_2)}e^{+(\lambda + \frac{\alpha_1 + \alpha_2}{2}) \slashed{r}}  (z_2  - \slashed{z_2})& \quad t > \lambda
\end{array}
\right. ,
\end{equation}

\begin{equation}
\begin{split}
\begin{pmatrix}
\left.
\begin{array}{ll}
a & b\\
c & d 
\end{array}
\right.
\end{pmatrix}^C
=
\begin{pmatrix}
b \\
d \\
-a \\
-c \\
\end{pmatrix}
\end{split} .
\end{equation}

Solution to Dirac Equation

\begin{equation}
\begin{split}
\chi_t = 
\left\{
\begin{array}{ll}
(e^{(t + \lambda)z_{1}}\Psi(-\lambda) - \frac{(t+\lambda)}{2 z_1}\rho(t)^{\dagger})^C & \quad t < -\lambda \\
\Psi(t)^C & \quad -\lambda < t < \lambda \\
(e^{(-t + \lambda)z_{2}}\Psi(\lambda) - \frac{(-t+\lambda)}{2 z_2}\rho(t)^{\dagger})^C & \quad t > \lambda
\end{array}
\right.  
\end{split} .
\end{equation}

\newpage

\end{document}